\documentclass[journal,a4paper,twoside]{IEEEtran}
\usepackage[utf8]{inputenc} % Allow direct UTF8-characters in the sources

\usepackage{units}

\usepackage{cite}
\ifCLASSINFOpdf \usepackage[pdftex]{graphicx}
\graphicspath{{fig/}}
\DeclareGraphicsExtensions{.pdf,.jpeg,.png} \else
\fi
\usepackage{tikz} \usetikzlibrary{shapes,arrows,calc,%
  decorations.markings,intersections}% automata,petri,backgrounds,
\definecolor{MyLightGreen}{RGB}{230,255,230}

\usepackage[cmex10]{amsmath}
\usepackage{algpseudocode}
\ifCLASSOPTIONcompsoc
\usepackage[caption=false,font=normalsize,labelfont=sf,textfont=sf]{subfig}
\else \usepackage[caption=false,font=footnotesize]{subfig} \fi

\usepackage{fixltx2e}

\newcommand{\thetitle}{Heater self-calibration technique for shape
  prediction of fiber tapers}

\newcommand{\theauthors}{Heidi~L.~Sørensen, Eugene~S.~Polzik, and
  Jürgen~Appel}

\usepackage{xcolor}%
\newcommand{\linkcolor}{magenta}%

\usepackage{hyperref} %% optional
\hypersetup{colorlinks=true, %
  linkcolor=\linkcolor, %
  citecolor=\linkcolor, %
  filecolor=\linkcolor, %
  urlcolor=\linkcolor, %
  pdftitle=\thetitle, pdfauthor=\theauthors %
} %

\usepackage{wasysym} %provides \permil

\renewcommand{\eqref}[1]{(\ref{#1})} %
\newcommand{\fig}[1]{Fig.~\ref{#1}} %

\newcommand{\zp}{z_+} \newcommand{\zm}{z_-} \newcommand{\zpm}{z_{\pm}}
\newcommand{\zpp}{\tilde{z}_{+}} \newcommand{\zmm}{\tilde{z}_{-}}

\newcommand{\vp}{v_+} \newcommand{\vm}{v_-} \newcommand{\vpm}{v_{\pm}}
\newcommand{\vinf}{v_{\infty}}

\newcommand{\lp}{l_+} \newcommand{\lm}{l_-} % pull lengths
\newcommand{\lpm}{l_{\pm}}

\newcommand{\dw}{d_{\text{w}}} % waist diameter
\newcommand{\An}{A_{\text{n}}} % Normalized cross-sectional area
\newcommand{\ud}{\,\mathrm{d}} %for integration
\newcommand{\diff}[2]{\frac{\mathrm{d}#1}{\mathrm{d}#2}}

\newcommand{\peerrev}[2]{\ifCLASSOPTIONpeerreview#1\else#2\fi}

\newcommand{\ieeeonly}[1]{#1}
\renewcommand{\ieeeonly}[1]{} % use this line to create the Arxiv version
\newcommand{\arxivonly}[1]{#1}% use this line to create the Arxiv version

\begin{document}
%
% paper title can use linebreaks \\
% within to get better formatting as desired Do not put math or
% special symbols in the title.
\title{\thetitle}

% author names and IEEE memberships use \thanks{} to gain access to
% the first footnote area a separate \thanks must be used for each
% paragraph as LaTeX2e's \thanks was not built to handle multiple
% paragraphs

\author{\theauthors% <-this % stops a space
  \thanks{\ieeeonly{Manuscript received MONTH DD, YYYY; revised MONTH
      DD, YYYY; accepted MONTH DD, YYYY. Date of publication MONTH DD,
      YYYY; date of current version MONTH DD, YYYY. This research has
      been supported by a grant from the US Army Research Office with
      funding from the DARPA QuASAR, EU project SIQS, and ERC grant
      INTERFACE.}}% <-this % stops a space
  \thanks{The authors are with the Niels Bohr Institute, University of
    Copenhagen, Blegdamsvej 17, DK-2100 K\o{}benhavn \O{}, Denmark
    (e-mail: \href{mailto:hls@nbi.dk}{hls@nbi.dk} or
    \href{mailto:jappel@nbi.dk}{jappel@nbi.dk}).}%
  % \thanks{Digital Object Identifier XXX}
}

% these prevent an unwanted space from occurring between the last
% author name and the end of the author line. i.e., if you had this:
% 
% \author{....lastname \thanks{...}  \thanks{...} }
% ^------------^------------^----Do not want these spaces!
%
% a space would be appended to the last name and could cause every
% name on that line to be shifted left slightly. This is one of those
% "LaTeX things". For instance, "\textbf{A} \textbf{B}" will typeset
% as "A B" not "AB". To get "AB" then you have to do:
% "\textbf{A}\textbf{B}" \thanks is no different in this regard, so
% shield the last } of each \thanks that ends a line with
% a % and do not let a space in before the next \thanks.
% Spaces after \IEEEmembership other than the last one are OK (and
% needed) as you are supposed to have spaces between the names. For
% what it is worth, this is a minor point as most people would not
% even notice if the said evil space somehow managed to creep in.

\ieeeonly{ \ifCLASSOPTIONpeerreview \markboth{Journal of Lightwave
    Technology,~Vol.~XX, No.~X, Month~20XX}%
  { \thetitle} \else \markboth{Journal of Lightwave
    Technology,~Vol.~XX, No.~X, Month~20XX}%
  {S\O{}rensen \MakeLowercase{\textit{et al.}}: \thetitle} \fi }
% The only time the second header will appear is for the odd numbered
% pages after the title page when using the twoside option.
% *** name in the headers of peer review papers.  *** You can use
% \ifCLASSOPTIONpeerreview for conditional compilation here if you
% desire.

% If you want to put a publisher's ID mark on the page you can do it
% like this: \IEEEpubid{0000--0000/00\$00.00~\copyright~2013 IEEE}
% Remember, if you use this you must call \IEEEpubidadjcol in the
% second column for its text to clear the IEEEpubid mark.

\ieeeonly{ \IEEEpubid{0000--0000/00\$00.00~\copyright~2013 IEEE} }

\maketitle

\ifdefined\svnid {\footnotesize
  \textcolor{green}{\svnFullRevision*{\svnrev} by
    \svnFullAuthor*{\svnauthor}, %\newline %
    Last changed date: \svndate } } \fi

% As a general rule, do not put math, special symbols or citations in
% the abstract or keywords.
\begin{abstract}\label{abstract}%
  In the production of tapered optical fibers, it is important to
  control the fiber shape according to application-dependent
  requirements and to ensure adiabatic tapers. Especially in the
  transition regions, the fiber shape depends on the heater
  properties. The axial viscosity profile of the fiber within the
  heater can, however, be hard to access and is therefore often
  approximated by assuming a uniform temperature distribution. We
  present a method for easy experimental calibration of the viscosity
  profile within the heater. This allows the determination of the
  resultant fiber shape for arbitrary pulling procedures, using only
  an additional camera and the fiber drawing setup itself. We find
  very good agreement between the modeled and measured fiber shape.
\end{abstract}

% { \hls{\svnFullRevision*{\svnrev} by \svnFullAuthor*{\svnauthor}
% \newline %
% Last changed date: \svndate } }


% page as needed: \ifCLASSOPTIONpeerreview
% \begin{center} \bfseries EDICS Category: 3-BBND \end{center} \fi
%
% For peerreview papers, this IEEEtran command inserts a page break
% and creates the second title. It will be ignored for other modes.

\IEEEpeerreviewmaketitle

% -------------------------------------------------------------------%
\section{Introduction}
% -------------------------------------------------------------------%
% The very first letter is a 2 line initial drop letter followed by
% the rest of the first word in caps.
% Here we have the typical use of a "T" for an initial drop letter and
% "HIS" in caps to complete the first word.  \IEEEPARstart{T}{his}
% demo file is intended to serve as a ``starter file'' for IEEE
% journal papers produced under \LaTeX\ using IEEEtran.cls version 1.8
% and later.

% \IEEEPARstart{W}{e} have seven pages in total to play on. There
% should not be any figures in the first column in the first page and
% preferable not on the first page at all. Figures should be placed in
% top of a page and never in the bottom. \hls{Remember to change from
% a4 to letter before submit and to change it to peer-review mode.}
	
\IEEEPARstart{D}{uring} the past decade the amazing properties of
tapered optical fibers (TOFs), such as low losses and the possibility
to concentrate large intensities in an evanescent light field, have
been extensively explored. They are now applied in various fields
ranging from optical bio-sensing to quantum optics
\cite{Balykin:2004bu,Xiao:2011cs,GarciaFernandez2011,lee2012oe,
  Park:2009gm,Bahl:2013eb,Sumetsky:2006gj}. In bio-sensing, TOFs with
biorecognition molecules in the evanescent field area
\cite{Fan:2008kn} enable the detection of specific target molecules;
in quantum optics, the high single-photon field strength near the
fiber surface allows for quantum light-atom interfaces employing only
a few atoms\cite{Alton:2010kl,Vetsch2010}.

A TOF is produced by heating a section of commercial optical fiber
while pulling its ends apart. Tailoring of the taper shape, such that
it fulfills the adiabaticity criteria\cite{Love:1986tz,Love:1991wq},
is important to avoid coupling light from the fundamental mode to
higher-order and radiation modes.

Traditionally, the tapered shape is modeled by assuming a uniform
viscosity profile of the heated fiber such that it is infinite outside
the heated section and finite inside
\cite{Birks1992,Baker:2011ws,Dewynne:1989uz}. From measured shapes we find that this simplified approach is not sufficient for TOFs produced in a microheater, as both the
waist and the shape of the tapers are not predicted correctly.  The
resulting shape of the TOF inherently depends on the strongly
temperature-dependent viscosity profile of the fiber induced by the
heating device.  While in flame-brushing fiber
processing~\cite{Kenny1991} the effective temperature distribution can
be relatively uniform, for heaters where this approximation fails, it
is necessary to include a fluid-dynamical description of the fiber
flow during the pulling procedure (given in
Sec.~\ref{sec:fibershape}).

Furthermore, we have confirmed that the fiber flow not only depends on
the applied boundary conditions of the pulling procedure, but also on
the momentary shape of the heat-softened fiber. This was also observed
in \cite{Pricking2010}, where the authors characterize their heater
and heuristically include its properties in a fiber-shape model.

In this work, we present a fluid-dynamics based model for the TOF
shape, taking into account both the axially varying viscosity profile
and the momentary shape-dependency of the axial velocity profile.

We provide a practical method to calibrate the parameters of the
heater via simple data analysis of the measured shape of a single TOF
made for this purpose. This calibration allows us to predict precisely
the shape of TOFs manufactured with arbitrary pulling procedures.

\IEEEpubidadjcol
% -------------------------------------------------------------------%
\section{Experimental setup}
% -------------------------------------------------------------------%
\subsection{Fiber-pulling rig}
\begin{figure}[!t]
  \centering
  \includegraphics[width=1.0\columnwidth]{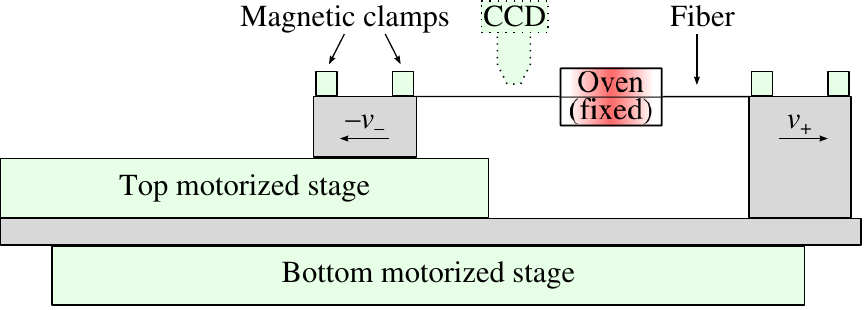}
  \caption{Schematic drawing of the fiber-pulling rig. The CCD camera
    is only inserted after the pulling procedure when the oven has
    been removed from the setup.}
  \label{fig:setup}
\end{figure}

% -------------------------------------------------------------------%
To produce a TOF we use a fiber-pulling rig consisting of two stacked
motorized linear translation stages and an NTT \textit{CMH-7019}
electric ceramic microheater (oven), as illustrated in
\fig{fig:setup}. The stacked solution of the stages provides improved
stability compared to stages placed in succession of each
other~\cite{Warken2008}. By only moving the bottom stage, the fiber
can easily be translated (without stretching it) with respect to the
stationary oven. During a pull, the bottom stage moves one end of the
fiber with velocity $\vp$, whereas the combined motion of the top and
bottom stages moves the other end with velocity $\vm$. In this paper
we restrict ourselves to cases where $\vp - \vm > 0$, i.e., where the
fiber is never compressed.

Driving the oven with electrical heating powers ranging from
\unit[97-103]{W}, we reproducibly observe identical TOF shapes. We
also confirm that the resultant shape only depends on the ratio of the
pull speeds $\vp/\vm$, rather than on their value, as long as the
speeds are sufficiently low that the fiber does not slip underneath
the magnetic clamps. This is a consequence of Newtonian fluid flow,
and (for a slow quasi-static pull) the fiber shape therefore only
depends on the pull lengths on either side of the oven.

% -------------------------------------------------------------------%
\subsection{Imaging}
\begin{figure}[!t]
  \centering \subfloat{ \centering%
    \includegraphics[width=1.0\columnwidth]{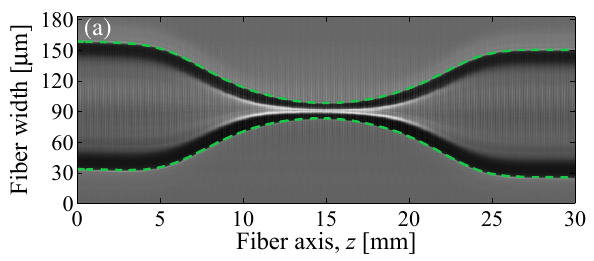}
    \label{fig:fiber}
  }\\
  \subfloat{ \centering%
    \includegraphics[width=1.0\columnwidth]{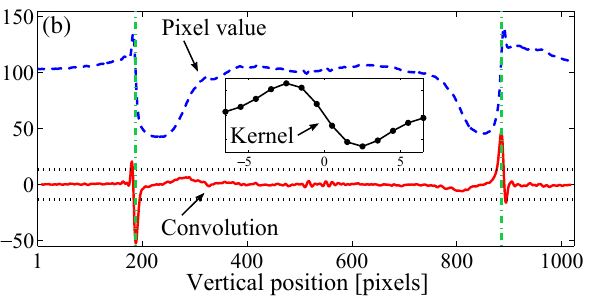}
    \label{fig:algorithm}
  }
  \caption{{(a)~300 joined CCD images of a TOF, symmetrically
      elongated by $l=\unit[15]{mm}$.  The waist is measured to
      $\dw=\unit[15]{\mu m}$. The aspect ratio is not to
      scale. (b)~Edge-detection algorithm. Dashed blue line: Pixel
      values along the image column at $z=\unit[5]{cm}$. Solid red
      line: Convolution. Dotted black lines: Threshold levels ($=\pm
      13$). Dashed-dotted green lines: Located edge positions. Inset:
      Edge-detection kernel.}}
\end{figure}

% -------------------------------------------------------------------%
To measure the TOF shape we image it with a CCD camera through a
$25\times$ microscope objective placed above the
fiber~(\fig{fig:setup}). The imaging is non-destructive, fast, and
\textit{in situ}: we obtain the full fiber shape by repeatedly
recording an image and translating the TOF with the bottom stage. The
individual images are joined, and a typical example of 300 merged
images is shown in~\fig{fig:fiber}. The green dashed curves indicate
the fiber edges found by an edge-detection algorithm: for each image
column, we calculate the convolution with a template kernel. We locate
the position of the edges by the outermost local minimum/maximum
values of the convolution that are significant enough to exceed a
threshold level, as indicated in~\fig{fig:algorithm}. Introducing this
threshold prevents detection errors caused by the narrow, bright
features close to the fiber axis; its value is set to $25\%$ of the
extremal convolution values found in the unstretched fiber. As
template kernel we use the derivative of a Gaussian with a width
chosen such that the kernel models the pixel values observed at the
edges of the unstretched fiber.

We estimate the precision of the diameter detection in two ways.
Determining the fiber diameter of an unstretched $\unit[125]{\mu m}$
fiber at $10^5$ image columns, we obtain a fiber width of
$\unit[694]{pixels}$ with a $\unit[1]{pixel}$ uncertainty in every
column. Additionally, after stretching the fiber by $\unit[15]{mm}$,
we observe only a relative change of the fiber volume $< 10^{-3}$
compared to the unstretched fiber. The dominating contribution to the
uncertainty is given by how well the diameter of the unstretched fiber
is known.

\begin{figure}
  \centering
  \includegraphics[width=1.0\columnwidth]{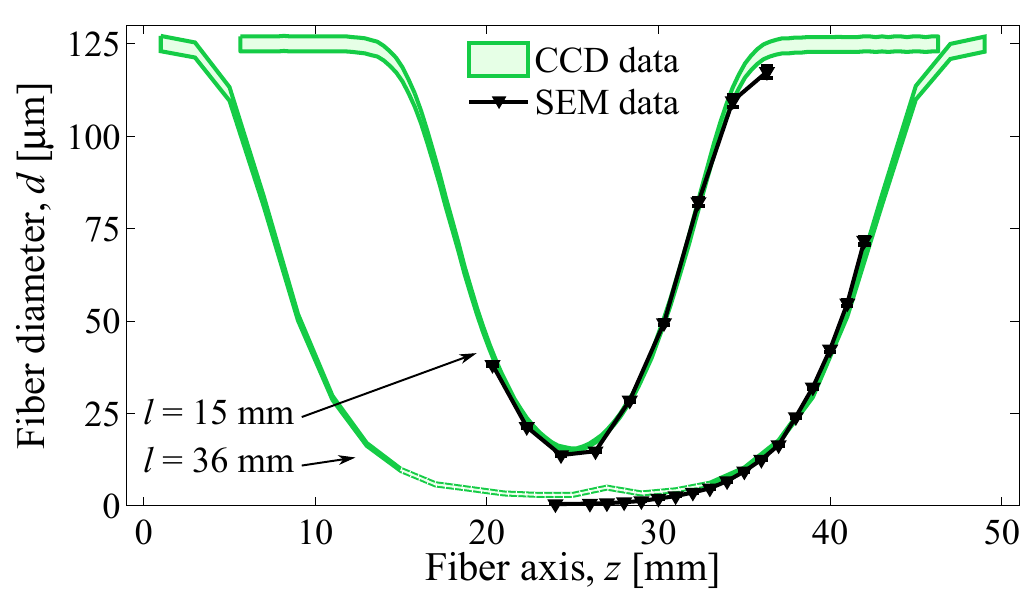}
  \caption{Comparison of fiber-diameter measuring methods. At several
    selected positions the fiber diameter is determined both using the
    CCD imaging and the SEM. For fiber diameters $<\unit[10]{\mu m}$
    diffraction effects limit the accuracy of the CCD method
    (indicated by the dashed lines), whereas the SEM method fails for
    large fibers due to space-charge buildup.}
  \label{fig:CCDSEM}
\end{figure}

Because the edge detection is limited by the optical imaging
resolution, diffraction effects, and the fiber bending out of the
focal plane, only TOFs with waist diameters larger than
$\approx\unit[10]{\mu m}$ can be measured. We confirm the validity of
our model also for thinner TOFs by additionally using a scanning
electron microscope (SEM) to measure the diameter at selected axial
positions, as shown in~\fig{fig:CCDSEM}. The SEM imaging is not
necessary for the presented calibration method, it is merely used for
verification.
%

% -------------------------------------------------------------------%
\section{Modeling the fiber shape}
% -------------------------------------------------------------------%

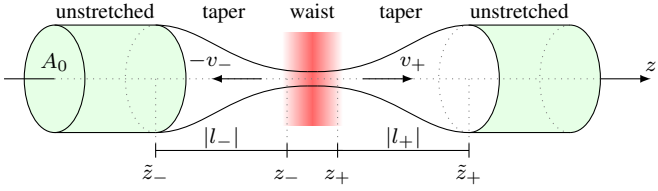
\begin{figure}[!t]
  \centering {\fontsize{8pt}{1em}\selectfont
    \begin{tikzpicture}[yscale=0.5,xscale=0.7,scale=0.95]
      % define line style:
      \tikzset{waist/.style={line width=0.4pt}}
      
      % top left transition coordinates
      \path coordinate (t0) at (-1,3) %
      coordinate (t1) at (1,3) %
      coordinate (t2) at (4,1.7)
      % coordinate (t2) at (4,2.3)
      coordinate (ct1) at (2.0,3) %
      coordinate (ct2) at (2.5,1.7);
      % coordinate (ct2) at (2.5,2.3)
		
      % top right transition coordinates
      \path coordinate (t3) at (4.2,1.7)
      % coordinate (t3) at (4.2,2.3)
      coordinate (t4) at (7.2,3) %
      coordinate (t5) at (9.2,3) %
      coordinate (ct3) at (5.7,1.7)
      % coordinate (ct3) at (5.7,2.3)
      coordinate (ct4) at (6.2,3);
		
      % bottom left transition coordinates
      \path coordinate (b0) at (-1,0) %
      coordinate (b1) at (1,0) %
      coordinate (b2) at (4,1.3) %
      % coordinate (b2) at (4,0.7)
      coordinate (cb1) at (2.0,0) %
      coordinate (cb2) at (2.5,1.3);
      % coordinate (cb2) at (2.5,0.7);
		
      % bottom right transition coordinates
      \path coordinate (b3) at (4.2,1.3) %
      % coordinate (b3) at (4.2,0.7)
      coordinate (b4) at (7.2,0) %
      coordinate (b5) at (9.2,0) %
      coordinate (cb3) at (5.7,1.3) %
      % coordinate (cb3) at (5.7,0.7)
      coordinate (cb4) at (6.2,0);
			
      % Color the unstrechted sections left
      \fill[MyLightGreen] (t0) rectangle (b1);%
      \fill[MyLightGreen] (1,1.5) ellipse(0.6 and 1.5);
			
      % right
      \fill[MyLightGreen] (t4) rectangle (b5);%
      \fill[white] (7.2,1.5) ellipse(0.6 and 1.5);
			
      % waist
      \fill[shading=horizontal,left color=white,right color=red!60]
      (3.5,0.2) rectangle (4.1,2.8);%
      \fill[shading=horizontal,left color=red!60,right color=white]
      (4.1,0.2) rectangle (4.7,2.8);

      % Draw and color left fiber end
      \filldraw[fill=MyLightGreen] (-1,1.5)node[above] {$A_0$}
      ellipse(0.6 and 1.5);%

      % Drawing the fiber edges left transition cross section
      \draw[] (1,0) arc (-90:90:0.6 and
      1.5);% right half of right ellipse
      \draw[dotted,gray] (1,0) arc (270:90:0.6 and
      1.5);% left half of the left ellipse

      % right transition cross section
      \draw (7.2,0) arc (-90:90:0.6 and
      1.5);% right half of right ellipse
      \draw[dotted,gray] (7.2,0) arc (270:90:0.6 and
      1.5);% left half of the left ellipse

      % right fiber end
      \fill[MyLightGreen] (9.2,1.5) ellipse(0.6 and 1.5);%
      \draw[] (9.2,0) arc (-90:90:0.6 and
      1.5);% right half of right ellipse
      \draw[dotted,gray] (9.2,0) arc (270:90:0.6 and
      1.5);% left half of the left ellipse

      % top edge
      \draw[] (t0) -- (t1);% left
      \draw[] (t1) .. controls (ct1) and (ct2) ..
      (t2);% left transition
      \draw[waist] (t2) -- (t3);% waist
      \draw[] (t3) .. controls (ct3) and (ct4) ..
      (t4);% right transition
      \draw[] (t4) -- (t5);% right

      % bottom edge
      \draw[] (b0) -- (b1);% bottom edge
      \draw[] (b1) .. controls (cb1) and (cb2) ..
      (b2);% left transition
      \draw[waist] (b2) -- (b3);% waist
      \draw[] (b3) .. controls (cb3) and (cb4) ..
      (b4);% right transition
      \draw[] (b4) -- (b5);% right

      % z axis and label
      \draw (-2,1.5) -- (-1,1.5);%
      \draw[dotted,gray] (-1,1.5) -- (9.8,1.5);%
      \draw[->,>=latex] (9.8,1.5) -- (10.8,1.5) node[above]{$z$};%

      % Length labels
      \draw (1,-0.5) -- node[above,yshift=-2]{$\vert\lm\vert$}
      (3.6,-0.5);%
      \draw (4.6,-0.5) -- node[above,yshift=-2]{$\vert\lp\vert$}
      (7.2,-0.5);%
      \draw (3.6,-0.5) -- (4.7,-0.5);%
      \draw[dotted,gray] (1,1.5) -- (1,-0.3);%
      \draw[dotted,gray] (3.6,1.5) -- (3.6,-0.3);%
      \draw[dotted,gray] (4.6,1.5) -- (4.6,-0.3);%
      \draw[dotted,gray] (7.2,1.5) -- (7.2,-0.3);%
      \draw (1,-0.3) -- (1,-0.7);%
      \draw (3.6,-0.3) -- (3.6,-0.7);%
      \draw (4.6,-0.3) -- (4.6,-0.7);%
      \draw (7.2,-0.3) -- (7.2,-0.7);%
      \node at (1,-1.3) [anchor=base] {$\zmm$};%
      \node at (3.6,-1.3) [anchor=base] {$\zm$};%
      \node at (4.6,-1.3) [anchor=base] {$\zp$};%
      \node at (7.2,-1.3) [anchor=base] {$\zpp$};%
	
      % velocities
      \draw[<-,>=latex] (2.1,1.5) node[above] {$-\vm$} -- (3.1,1.5);
      \draw[->,>=latex] (5.1,1.5) -- (6.1,1.5) node[above]{$\vp$};

      % Fiber sections terminology
      \node at (0,3.2) [anchor=base] {unstretched};%
      \node at (2.35,3.2)[anchor=base] {taper};%
      \node at (4.1,3.2)[anchor=base] {waist};%
      \node at (5.85,3.2)[anchor=base] {taper};%
      \node at (8.2,3.2) [anchor=base] {unstretched};%
    \end{tikzpicture}
  }
  \caption{Boundary conditions during the fiber pulling
    procedure. $l_{\pm} = v_{\pm} t = \tilde{z}_{\pm}-z_{\pm}$ are the
    elongated lengths of the fiber on either side of the heated
    section represented by the red area in the center.}
  \label{fig:boundary}
\end{figure}

We consider an optical fiber with position- and time-dependent
cross-sectional area $A(z,t)$ passing through a heater, as illustrated
in~\fig{fig:boundary}.

% -------------------------------------------------------------------%
\subsection{Boundary conditions}
% -------------------------------------------------------------------%
The axial fiber flow $v(z,t)$, at position $z$ and time $t$ for
regions on either side of the heated section corresponds to the speed
of the respective fiber holders:%
\begin{subequations}
  \label{eq:vBoundary}
  \begin{align}
    v(z,t) &= \vm, \quad \text{for } z < \zm\,, \\
    v(z,t) &= \vp, \quad \text{for } z > \zp\,,
  \end{align}
\end{subequations}
where the sign of $\vpm$ follows the direction of the pull. Inside the
heated zone, $v(z,t)$ is described by an unknown function that depends
on the pull speeds, the momentary fiber shape, and the axial viscosity
distribution of the fiber, resulting from the axial temperature
profile of the heater (neglecting any transverse variation).
The boundary between each taper and the unstretched fiber is denoted
by $\tilde{z}_\pm$. Outside the tapered sections ($z<\zmm$ or
$z>\zpp$), the cross-sectional area corresponds to that of the
initially uniform fiber, $A(z,t)=A_0$. We introduce the following
abbreviations:
\begin{subequations}
  \begin{align}
    \vinf &\equiv \vp-\vm\, & & \text{denotes the stretching speed}, \\
    % l &= \lp - \lm = \vinf
    % t\,, \label{eq:l}\\
    \An(z,t) &\equiv \frac{A(z,t)}{A_0}\, & & \text{the normalized
      cross-sectional area}.
    % a &= \frac{A_0}{A(z,t)}\,,
  \end{align}
\end{subequations}
%
% -------------------------------------------------------------------%
\subsection{Fiber shape \label{sec:fibershape}}
% -------------------------------------------------------------------%
The evolution of the fiber shape during the tapering procedure can be
described by two coupled differential equations for the normalized
cross-sectional area $\An(z,t)$ and the axial velocity profile of the
fiber $v(z,t)$~\cite{Geyling:1976uz,Dewynne:1989uz}. The continuity
equation
\begin{equation}
  \frac{\partial}{\partial t} \An(z,t) + 
  \frac{\partial}{\partial z} \Big( \An(z,t) \, v(z,t) \Big) = 0 \,
  \label{eq:continuity}
\end{equation}
governs mass conservation, and a simplified equation describes axial
momentum conservation:
\begin{equation}
  \frac{\partial}{\partial z} 
  \Big( \eta(z)	\An(z,t) \frac{\partial}{\partial z} v(z,t) \Big) =0\,,
  \label{eq:vdiff}
\end{equation}
where $\eta(z)$ is the axial viscosity of the fiber
fluid. Equation~\eqref{eq:vdiff} is derived by solving the
Navier-Stokes equations for an axisymmetric incompressible Newtonian
fluid in the limit of Stokes flow, neglecting body forces (such as
gravity, which is negligible compared to viscous forces), and by
Taylor-expanding the equations to lowest order in the radial variable
\cite{Eggers1994}.  The fiber is thin, and its heat conductivity is
poor compared to that of the much bigger surrounding oven. Therefore
the temperature along the fiber (and hence $\eta(z)$) is a function of
the axial position within the oven alone. Additionally, we ensure that
each mass element of the fiber is in thermal equilibrium with the
surroundings by asserting slow motion of the fiber.

In order to solve~\eqref{eq:continuity} and~\eqref{eq:vdiff}
numerically, it is necessary to know $\eta(z)$. Often, this is simply
approximated by a uniform distribution such that it is infinite
outside the heated region of the fiber and finite and constant inside
\cite{Dewynne:1989uz,Birks1992,Baker:2011ws}. Xue \emph{et
  al.}~\cite{Xue:2007gu} measure the temperature distribution of their
heater and use the Arrhenius model for the viscosity dependence on the
temperature to indirectly deduce $\eta(z)$. Pricking \emph{et
  al.}~\cite{Pricking2010} heuristically model $\eta(z)$ by a
flattened Gaussian profile. In the following we show how $\eta(z)$
instead can be easily inferred experimentally by measuring the
resultant fiber shape after a short symmetric ($-\vm=\vp$) pull. We
thereby avoid cumbersome temperature-viscosity calibrations and
measurements of the temperature profile inside the heater.

% -------------------------------------------------------------------%
\subsection{Fiber fluidity}
% -------------------------------------------------------------------%
Since the TOF shape depends on the ratio of the velocities, and not on
the individual velocities, only a dependence on the pull lengths $\lpm
= \vpm t$ remains. It is therefore more convenient and intuitive to
express the following equations in terms of the total elongation
length
\begin{equation}
  l=\lp-\lm = \vinf t
\end{equation}
instead of time, such that $v(z,t) \rightarrow v(z,l)$ and $\An(z,t)
\rightarrow \An(z,l)$.

Integrating~\eqref{eq:vdiff} over $z$ yields
\begin{equation}
  \eta(z)	\An(z,l) \frac{\partial}{\partial z} v(z,l) = C(l);
  \label{eq:intvdiff}
\end{equation}
the integration constant $C(l)$ does not depend on $z$. We
solve~\eqref{eq:intvdiff} for $\frac{\partial}{\partial z} v(z,l)$ and
integrate over $z$ again, starting at an arbitrary position $z_0$, and
obtain
\begin{equation}
  \label{eq:v}
  v(z,l) = v(z_0,l) + \vinf \cdot 
  \frac{ \int_{z_0}^{z} \frac{\tau(\zeta)}{\An(\zeta,l)} \ud \zeta } 
  { \int_{\zm}^{\zp} \frac{\tau(\zeta)}{\An(\zeta,l)} \ud\zeta
  }\,,
\end{equation}
for the axial velocity profile. The integration constant
\begin{equation}
  C(l)=\frac{\vinf}{\int_{z_-}^{z_+} \frac{1}{\eta(\zeta) \An(\zeta,l)} \ud\zeta}
\end{equation}
has been fixed by requiring continuity at the boundaries
$v(\zpm,l)=\vpm$; we also introduced the normalized fiber fluidity:
\begin{equation}
  \tau(z)=\frac{\frac{1}{\eta(z)} }
  { \int_{-\infty}^{\infty} \frac{1}{\eta(\zeta)} \ud \zeta}\,.
\end{equation}
Please note that $\tau(z)$ only differs from zero inside the heated
section bounded by $\zpm$.

% -------------------------------------------------------------------%
\subsection{Short pull}
% -------------------------------------------------------------------%
In the following, we consider a symmetric pull where the fiber
elongation length $l$ is much smaller than the heated section. In this
limit, the spatial variation of the normalized fiber cross-sectional
area $\An(z,l)$ over regions with non-zero $\tau(z)$ can be neglected
and~\eqref{eq:v}, describing the axial velocity profile, simplifies
significantly. If $z_0$ is chosen outside the heat-softened section,
such that $v(z_0,l)$ is constant, we find
\begin{equation}
  v(z,l)\approx v(z) = v(z_0) + \vinf \int_{z_0}^{z} \tau(\zeta) \ud \zeta
\end{equation}
to be constant in $l$ during the whole pulling process. From this, the
fiber fluidity can be readily approximated by
\begin{equation}
  \label{eq:tau}
  \tau(z) \approx \diff{}{z} \frac{v(z,l)}{\vinf}\,. 
\end{equation}
Since the axial velocity profile is now independent of the elongation
length, the continuity equation~\eqref{eq:continuity} can be solved
analytically to yield an explicit form for the normalized fiber
cross-sectional area:
\begin{subequations}
  \label{eq:Aq}
  \begin{align}
    \An(z,l) &= \frac{\partial}{\partial z} \Bigl( q^{-1}\bigl( q(z) - l \bigr) \Bigr) \label{eq:A} \\
    \text{with } q(z) & \equiv \int_{z_*}^{z} \frac{\vinf}{v(\zeta)}
    \ud\zeta \,,
    \label{eq:q}
  \end{align}
\end{subequations}
as can be directly verified by differentiation, i.e., by inserting
\eqref{eq:A} and \eqref{eq:q} into
\eqref{eq:continuity}. $q^{-1}(\cdot)$ denotes the inverse function of
$q(z)$, and $z_*$ is an arbitrarily chosen position. We integrate both
sides of~\eqref{eq:A} from $\tilde z_\pm$ and define a new variable
\begin{subequations}
  \label{eq:y}
  \begin{align}
    y(z,l) &\equiv  q^{-1}\big( q(z) - l \big) \label{eq:ya} \text{, so that} \\
    y(z,l) &= \int_{\tilde z_\pm}^{z} \An(\zeta,l) \,\ud\zeta \, +
    z_\pm.
    \label{eq:yb}
  \end{align}
\end{subequations}
The second term in~\eqref{eq:yb} follows from choosing $z_*=\zpm$
in~\eqref{eq:q}. The expression $z-y(z,l)$ can be interpreted as the
distance that a fiber volume element at position $z$ has moved during
the pulling process.

We apply $q(\cdot)$ to both sides of~\eqref{eq:ya} and differentiate
with respect to $z$. Using~\eqref{eq:yb} to express $\frac{\partial
  y}{\partial z}$ and~\eqref{eq:q} to express $\diff{q}{z}$ in the
result, we obtain a recursion formula for the axial velocity profile
of the fiber:
\begin{equation}
  v\bigl( y(z,l) \bigr) = \An(z,l)\, v(z)\,.  \label{eq:vRec}
\end{equation}
Both $\An(z,l)$ and $y(z,l)$ are known from the fiber shape
measurements (the latter via ~\eqref{eq:yb}). On the left side of the
oven $y(z,l)>z$. Starting from $z = \tilde z_-$, using
\eqref{eq:vRec}, we can now calculate $v(y(z,l))$ from $v(z)$, which
lies further to the left, until $y(z,l)$ approaches $z$. The same can
be done from the other side starting from $z=\tilde z_+$, since there
$y(z,l)<z$.

The following pseudo-code illustrates the algorithm for calculating
$v(z)$ in the interval $[\zmm,\zpp]$ with a step size $\Delta z>0$:
\begin{algorithmic}[1]
  \State vz$_{\text{table}} \gets \big\{(v_-,\tilde z_-), \,
  (v_+,\tilde z_+)\big\};$%
  \State $z \gets \tilde z_-; \quad v \gets v_-; \quad y \gets \tilde
  z_- - l_-;$%
  \While{$ y > z$}%
  % \Comment from left side%
  \State insert $\big(v\cdot \An(z),y\big)$ into
  $\text{vz}_{\text{table}};$%
  \State $z \gets z + \Delta z;$%
  \State $y \gets y + \Delta z\cdot A_{\text{n}}(z);$%
  \State $v \gets \texttt{interpolate}(\text{vz}_{\text{table}},z);$
  \EndWhile; \State $z \gets \tilde z_+; \quad v \gets v_+; \quad y
  \gets \tilde z_+ - l_+;$%
  \While{$ y< z $}%
  % \Comment from right side%
  \State insert $\big(v\cdot \An(z),y\big)$ into
  $\text{vz}_{\text{table}};$%
  \State $z \gets z - \Delta z;$%
  \State $y \gets y - \Delta z\cdot A_{\text{n}}(z);$%
  \State $v \gets \texttt{interpolate}(\text{vz}_{\text{table}},z);$%
  \EndWhile;
\end{algorithmic}
In this way the complete velocity profile $v(z)$ is contained in
$\text{vz}_\text{table}$, and by~\eqref{eq:tau} the fiber fluidity can
be calculated.

% -------------------------------------------------------------------%
\section{Results}
% -------------------------------------------------------------------%

% -------------------------------------------------------------------%
\subsection{Calibration}
% -------------------------------------------------------------------%
%
\begin{figure}[!t]
  \subfloat{%
    \centering%
    \includegraphics[trim=0 87 0 0,clip=true,%
    keepaspectratio,width=1.0\columnwidth]{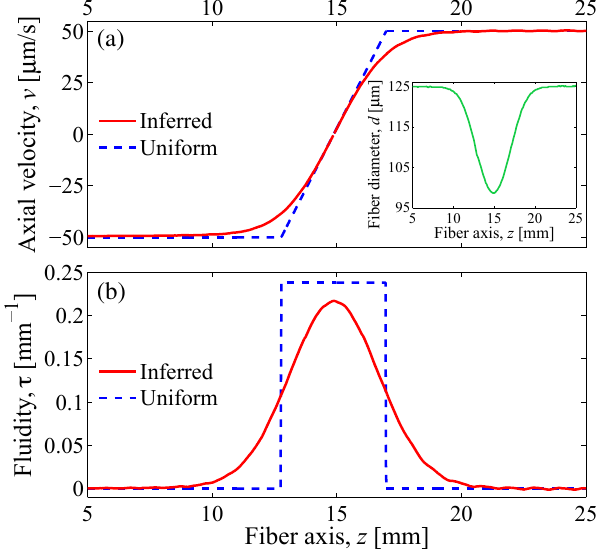}%
    \label{fig:vtaua}%
  }\\
  \subfloat{%
    \centering%
    \includegraphics[trim=0 2 0 76, clip=true,%
    keepaspectratio,width=1.0\columnwidth]{soere3}%
    \label{fig:vtaub}%
  }
  \caption{(a)~Axial velocity profile. Inset: measured fiber diameter
    of an $l=\unit[2]{mm}$ stretched fiber used to infer
    $v(z,l)$. (b)~Axial fluidity profile, inferred by
    applying~\eqref{eq:tau} to (a). Red solid lines depict data
    inferred with the presented algorithm. Blue dashed lines depict
    data corresponding to a uniform fluidity profile with
    $L_0=\unit[4.2]{mm}$, which results in an identical waist
    diameter.}
  \label{fig:vtau}
\end{figure}
To the measured shape $d(z,l)=2\sqrt{A(z,l)/\pi}$ of a TOF which was
elongated symmetrically by $l=\unit[2]{mm}$ with speeds $\vpm=\pm
\unit[50]{\mu m/s}$, we apply the recursion formula~\eqref{eq:vRec} to
infer the axial velocity profile $v(z,l)$ shown in~\fig{fig:vtaua}.
Also depicted is a simplifying model which was introduced in the
seminal paper by Birks and Li~\cite{Birks1992}. It is commonly used to
describe flame-brushing fiber processing~\cite{Kenny1991}, and it
approximates $v(z)$ inside the heated section by interpolating
linearly between the exterior pull velocities $v_\pm$ over an
``effective hot-zone length'' $L_0=\left. v_\infty/\frac{\ud v}{\ud
    z}\right|_{v=0}$.

The effective hot-zone length $L_0$ can be found from the waist
diameter $\dw$ using Birks' and Li's formula:
\begin{equation}
  \label{eq:dwBirks}
  \dw(l)=d_0 \exp{\bigg(-\frac{l}{2L_0}\bigg)}\,.
\end{equation}

To calibrate $\eta(z)$, we use a TOF with an initial diameter
$d_0=\unit[125]{\mu m}$ and a final waist diameter $\dw=\unit[98]{\mu
  m}$, which results in $L_0=\unit[4.2]{mm}$.

The curves for~$v(z)$ in~\fig{fig:vtaua} agree in value and slope at
the oven center and at the ends by construction but they deviate
substantially at the edges of the heated section. The difference is
even more pronounced in $\tau(z)$, which is depicted
in~\fig{fig:vtaub}. This strongly suggests that the assumption of a
uniform temperature distribution does not describe our setup.

% -------------------------------------------------------------------%
\subsection{Modeling a symmetric pull}
% -------------------------------------------------------------------%
Given the inferred fiber fluidity $\tau(z)$ we numerically solve the
system of equations~\eqref{eq:continuity} and~\eqref{eq:v}, using the
\texttt{MATLAB} function \texttt{ode45} with a relative error
tolerance of $10^{-6}$. For thin TOFs with diameters below
$\approx\unit[1]{\mu m}$, numerical instabilities can occur, which
necessitates decreasing the relative and absolute error tolerances
further. Alternatively, by adding a term $D\frac{\partial^2 \An(z,t)}
{\partial z^2}$ to the right-hand side of \eqref{eq:continuity}, using
a small ``diffusion coefficient'' $D$ such that $2\sqrt{D l/\vinf} \ll
L_0$, we can effectively eliminate the numerical stiffness of the
problem without introducing a significant change to the solution.

\begin{figure}[!t]
  \centering%
  \includegraphics[keepaspectratio,width=1.0\columnwidth]{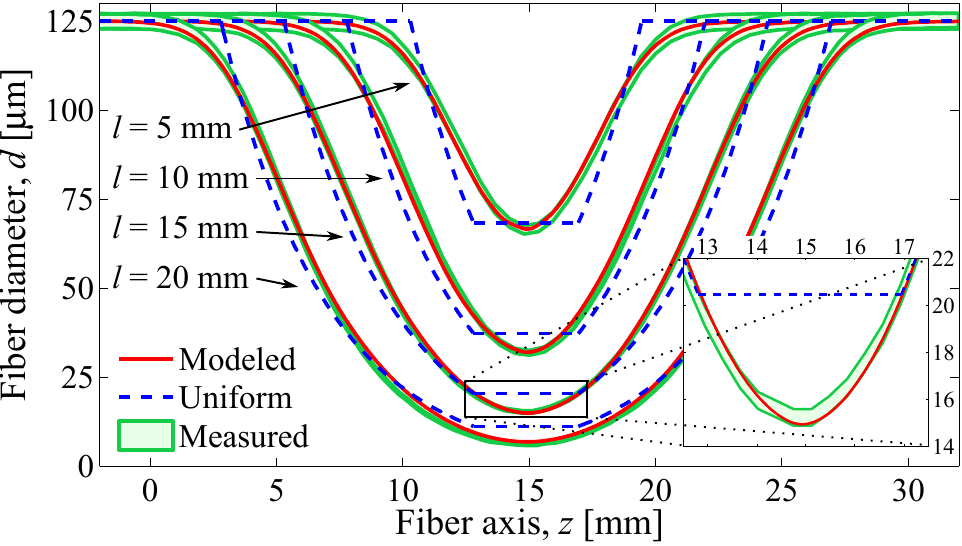}%
  \caption{Fiber diameter of four different symmetrically pulled
    fibers with various elongation lengths $l$. Solid red line: Model.
    Green lines enclosing shaded area: Fiber diameter as measured with
    CCD camera with measurement uncertainty. Dashed blue line: Fiber
    shape prediction using an $L_0=\unit[4.2]{mm}$ uniform fluidity
    profile. The inset shows a zoom of the waist of the fiber
    stretched by $l=\unit[15]{mm}$.}
  \label{fig:sym}
\end{figure}

In~\fig{fig:sym}, we present the modeling of four symmetrically
stretched fibers, which were elongated by $l=\unit[5,10,15,20]{mm}$
with speeds $\vpm=\pm \unit[50]{\mu m/s}$. We observe very good
agreement between the measured and modeled diameter with only a $1\%$
discrepancy at the waist of the $l=\unit[5]{mm}$ and $l=\unit[10]{mm}$
streched fibers, and $2\%$ for the $l=\unit[15]{mm}$ fiber. For the
longer $l=\unit[20]{mm}$ stretched fiber the discrepancy is 13\%;
however, here the waist is so thin, $\dw \approx \unit[6]{\mu m}$,
that the CCD imaging starts to fail.

For reference, we also show the predicted TOF shape using a uniform
profile for the fluidity with $L_0=\unit[4.2]{mm}$, which (by
definition) predicts the waist correctly for $l=\unit[2]{mm}$. For
this $\tau(z)$ it is evident that the waist size is increasingly
overestimated for longer pull lengths. As can also be observed by
numerically solving~\eqref{eq:vdiff}, this implies that the effective
hot-zone length $L_0$~\eqref{eq:dwBirks} of the fiber shrinks during
the pull (i.e., for smaller fiber diameters) in agreement with similar
observations made in~\cite{Pricking2010}. This shape-dependency makes
it impossible to predict the waist for arbitrary pull lengths using a
constant-width box-profile for the fluidity, as it fails to reproduce
qualitative features of the TOF shape. Especially the prediction of a
homogeneous waist with length $L_0$ is absent in the data.  This
necessitates non-symmetric pulling procedures for producing TOFs with
long homogeneous waists.

\begin{figure}[!t]
  \centering%
  \includegraphics[keepaspectratio,width=1.0\columnwidth]{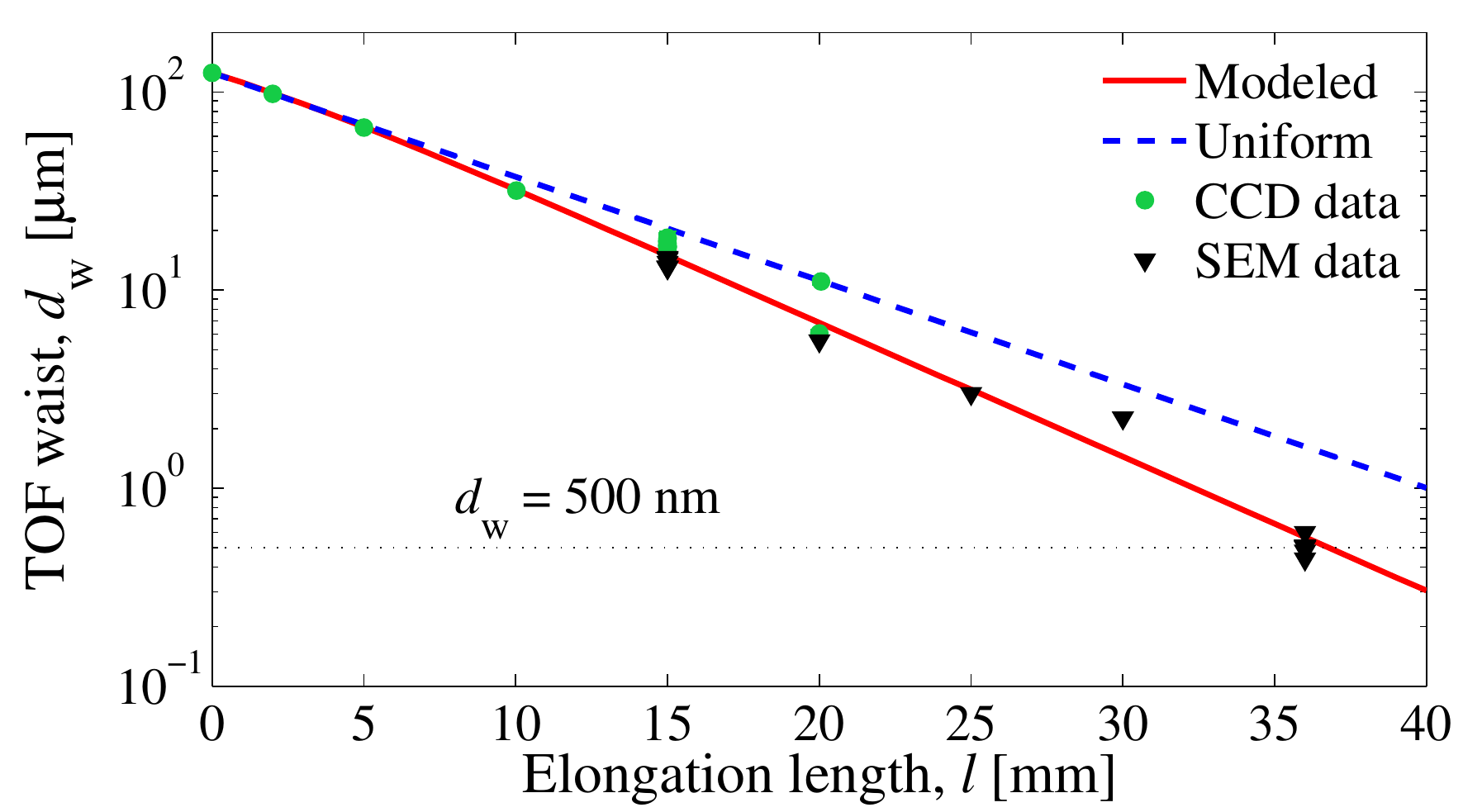}%
  \caption{TOF waist of symmetrically elongated fibers. Note the log
    scale on the y-axis.}
  \label{fig:modeledWaist}
\end{figure}

In~\fig{fig:modeledWaist}, for symmetric pulls, we compare the
predicted fiber waist diameter resulting from our calibration method
with experimental data and the simplified prediction
(\ref{eq:dwBirks}). Whereas the latter overestimates the waist for
longer pull lengths, our simulations display good agreement with the
data even for very thin TOFs, where the initial diameter has been
reduced by a factor of 250 from $\unit[125]{\mu m}$ to
about~$\unit[500]{nm}$.
In trying to fit (\ref{eq:dwBirks}) to the data shown in
\fig{fig:modeledWaist} by determining an \emph{effective}
$L_0$~\cite{Kenny1991,Ward2006,Ding2010}, one would compromise on the
predicted corresponding shape of the tapers instead.

% -------------------------------------------------------------------%
\subsection{Modeling an asymmetric pull}
% -------------------------------------------------------------------%
The fiber shape model presented here is not only restricted to
symmetric pulls, where $-\vm=\vp$, but can be applied to any
combination of pull speeds. This is extremely useful as it makes it
possible to test various pulling procedures without actually
performing them.

\begin{figure}[!t]
  \centering
  \includegraphics[keepaspectratio,width=1.0\columnwidth]{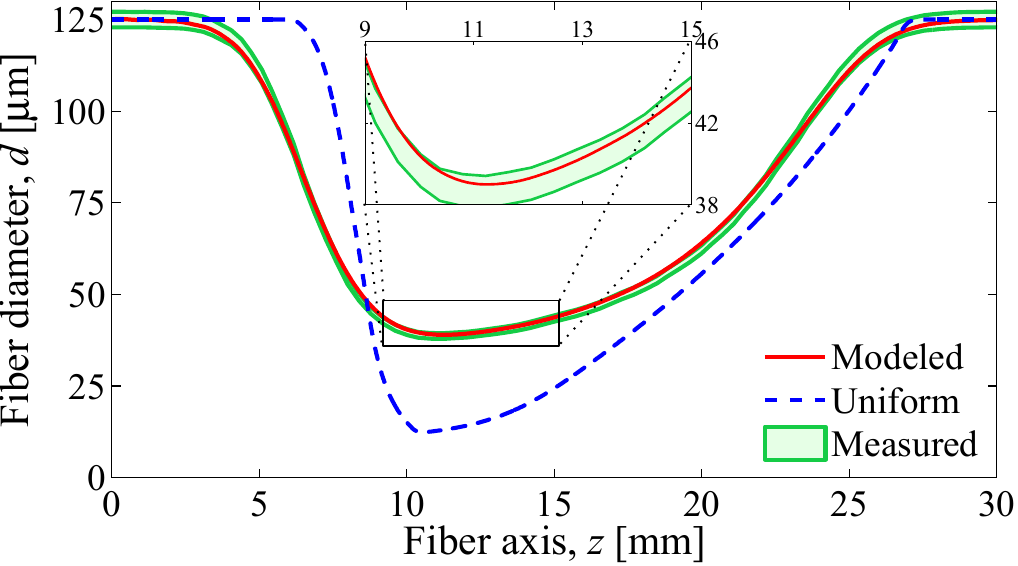}%
  \caption{Shape of a TOF, asymmetrically elongated by
    $l=\unit[15]{mm}$. Red solid line: Predicted fiber shape using the
    inferred fluidity profile depicted in \fig{fig:vtau}. Green lines
    enclosing shaded area: CCD measured diameter. Blue dashed line:
    Solution to~\eqref{eq:continuity} and~\eqref{eq:v} using a uniform
    fluidity profile with~$L_0=\unit[4.2]{mm}$. The inset shows a zoom
    of the fiber waist.}
  \label{fig:asym}
\end{figure}

Here we show the extreme situation where the two fiber ends are moved
in the same direction such that the fiber is being pushed into the
oven from one side while being pulled out on the other side with a
greater speed, i.e., $0<\vm<\vp$. The measured and modeled diameter of
such an asymmetrically pulled TOF is shown in~\fig{fig:asym}. Here, an
elongation of~\unit[15]{mm} is obtained by push and pull speeds
$\vm=\unit[10]{\mu m/s}$ and $\vp=\unit[100]{\mu m/s}$. The modeled
curve predicts the data very closely with only a 3\% discrepancy at
the waist and well within the uncertainty of the CCD data.

We demonstrate that especially in a situation where the axial fiber
diameter changes strongly within the heated zone, accurate modeling of
the viscosity profile leads to a significant improvement of the fiber
shape prediction.

\section{Conclusion}
% -------------------------------------------------------------------%
We present a general fiber shape model applicable to arbitrary pulling
procedures. Crucial for the model is the fluidity profile of the fiber
provided by the heater. We show how this can be inferred
experimentally using the fiber-pulling apparatus itself, assuming only
a temperature profile inside the oven that does not depend on the
fiber shape. The experimental calibration of the heater properties
allows a precise numerical prediction of the fiber shape. We thus
expect our method to facilitate the design of pulling procedures for a
wide range of applications requiring precision control of the fiber
shape. These include manufacturing of fiber-couplers and tapered
optical nanofibers for atom traps and nano-photonics. Whereas the
method presented here only allows one to predict precisely the shape
of the TOF for a given pulling procedure, the inverse problem is of
central interest for production of TOFs. In the case that a uniform
fluidity profile describes the physics sufficiently well, algorithms
to approximate the intended fiber shape do
exist\cite{Baker:2011ws}. Using such a solution as a starting point,
the proposed method could be used to solve a variational optimization
problem, adapting trial pulling procedures to regimes where other
algorithms fail. We are currently working on implementing this
approach.

% \appendix[Proof of the Zonklar Equations]
% or
% \appendix % for no appendix heading
% do not use \section anymore after \appendix, only \section* is
% possibly needed
	
% use appendices with more than one appendix then use \section to
% start each appendix you must declare a \section before using any
% \subsection or using \label (\appendices by itself starts a section
% numbered zero.)

% \appendices

% you can choose not to have a title for an appendix if you want by
% leaving the argument blank
% \section{}

\section*{Acknowledgment}
The authors would like to thank S.L.~Christensen and J.H.~Müller for
help in setting up the experiment and for inspiring discussions
concerning the imaging setup, and A.~Fabricant for help in editing the
manuscript.

% Can use something like this to put references on a page by
% themselves when using endfloat and the captionsoff option.
\ifCLASSOPTIONcaptionsoff \newpage \fi

% trigger a \newpage just before the given reference number - used to
% balance the columns on the last page adjust value as needed - may
% need to be readjusted if the document is modified later
% \IEEEtriggeratref{8} The "triggered" command can be changed if
% desired: \IEEEtriggercmd{\enlargethispage{-5in}}

% \ieeeonly{ \IEEEtriggeratref{12} }
\arxivonly{ \IEEEtriggeratref{8} } 

% references section can use a bibliography generated by BibTeX as a
% .bbl file BibTeX documentation can be easily obtained at:
% http://www.ctan.org/tex-archive/biblio/bibtex/contrib/doc/ The
% IEEEtran BibTeX style support page is at:
\bibliographystyle{IEEEtran} %
\bibliography{bibliography.bib} %
% argument is your BibTeX string definitions and bibliography
% database(s) \bibliography{IEEEabrv,../bib/paper}
%
% <OR> manually copy in the resultant .bbl file set second argument
% of \begin to the number of references (used to reserve space for the
%   reference number labels box)
%   \begin{thebibliography}{1}
%
%   \bibitem{IEEEhowto:kopka} H.~Kopka and P.~W. Daly, \emph{A Guide
%     to \LaTeX}, 3rd~ed.\hskip 1em plus 0.5em minus 0.4em\relax
%     Harlow, England: Addison-Wesley, 1999.
%
%   \end{thebibliography}

%   biography section
% 
%   If you have an EPS/PDF photo (graphicx package needed) extra
%   braces are needed around the contents of the optional argument to
%   biography to prevent the LaTeX parser from getting confused when
%   it sees the complicated
%   \includegraphics command within an optional argument. (You could
%   create your own custom macro containing the \includegraphics
%   command to make things simpler here.)
%   \begin{IEEEbiography}[{\includegraphics[width=1in,height=1.25in,clip,keepaspectratio]{mshell}}]{Michael
%     Shell}
%     or if you just want to reserve a space for a photo:

%     \begin{IEEEbiography}{Michael Shell}
%       Biography text here.
%     \end{IEEEbiography}

\ieeeonly{ \peerrev{}{
    \newcommand{\photo}[1]{\includegraphics[width=1in,height=1.25in,clip,keepaspectratio]{#1}}

    \vspace{-0.3cm}
    \begin{IEEEbiography}[\photo{soere}]{Heidi L. S\o{}rensen}
      received the Master’s degree in physics in 2013 from the Niels
      Bohr Institute, University of Copenhagen, Denmark. She is
      currently working towards the Ph.D. degree in the same
      department implementing a light-atom quantum interface using a
      tapered optical fiber.
      \label{bio:hls}
    \end{IEEEbiography}

    % insert where needed to balance the two columns on the last page
    % with biographies \newpage
    \vspace{-0.3cm}
    \begin{IEEEbiography}[\photo{polzi}]{Eugene Polzik}
      is a professor of physics and director of the Center for Quantum
      Optics – QUANTOP at the Niels Bohr Institute in Copenhagen. His
      research interests are centered around quantum interface between
      light and matter, quantum communication and quantum limited
      sensing. Dr. Polzik has published more than 130 papers in
      refereed journals and presented more than 150 invited and
      plenary talks at major international conferences. He is a Member
      of the Royal Danish Academy of Science, a Fellow of American
      Physical Society, a Fellow of Optical Society of America, and a
      Fellow of the Institute of Physics.
      \label{bio:ep}
    \end{IEEEbiography}
    \vspace{-0.3cm}
    \begin{IEEEbiography}[\photo{appel}]{Jürgen Appel}
      received the Ph.D. degree in physics from the University of
      Calgary, Canada in 2007. He is currently Associate Professor for
      quantum optics at the Niels Bohr Institute, University of
      Copenhagen, Denmark. His research interests include quantum
      noise limited measurements, entanglement and light-atom quantum
      interfaces. He is currently investigating the use of nano-fibers
      to engineer scalable, strong coupling to few quantum emitters.
      \label{bio:ja}
    \end{IEEEbiography}

    % You can push biographies down or up by placing a \vfill before
    % or
    % after them. The appropriate use of \vfill depends on what kind
    % of
    % text is on the last page and whether or not the columns are
    % being
    % equalized.
    \vfill } }
% Can be used to pull up biographies so that the bottom of the last
% one is flush with the other column.  \enlargethispage{-5in}

\end{document}